\documentclass[reprint,superscriptaddress,amsmath,amssymb,aps,prl]{revtex4-2}

\usepackage[english]{babel}
\usepackage[utf8]{inputenc}
\usepackage{graphicx}
\usepackage{hyperref}
\usepackage{xcolor}
\usepackage{amsmath}
\usepackage{amsfonts}
\usepackage{amssymb}
\usepackage{dsfont}
\usepackage{siunitx}
\usepackage{xcolor,import}
\usepackage{transparent}
\usepackage[T1]{fontenc}

\renewcommand{\selectlanguage}[1]{}

\begin{document}

\title{Superradiance of Strongly Interacting Dipolar Excitons in Moir\'e Quantum Materials}

\author{Jan Kumlin}%
\email{jan.kumlin@tuwien.ac.at}
\affiliation{Institute for Theoretical Physics, TU Wien, Wiedner Hauptstraße 8-10/136, A-1040 Vienna, Austria}%

\author{Ajit Srivastava}%
\affiliation{Department of Physics, Emory University, Atlanta, Georgia, USA}%
\affiliation{Department of Quantum Matter Physics, University of Geneva, Geneva, Switzerland}%

\author{Thomas Pohl}%
\affiliation{Institute for Theoretical Physics, TU Wien, Wiedner Hauptstraße 8-10/136, A-1040 Vienna, Austria}%

\date{\today}

\begin{abstract}
    Moir\'e lattices created in two-dimensional heterostructures exhibit rich many-body physics of interacting electrons and excitons and, at the same time, suggest promising optoelectronic applications. Here, we study the cooperative radiance of moir\'e excitons that is demonstrated to emerge from the deep subwavelength nature of the moir\'e lattice and the strong excitonic on-site interaction. In particular, we show that the static dipole-dipole interaction between interlayer excitons can  strongly affect their cooperative optical properties, suppressing superradiance of disordered states while enhancing superradiance of ordered phases of moir\'e excitons. Moreover, we show that doping permits direct control of optical cooperativity, e.g., by generating superradiant dynamics of otherwise subradiant states of excitons. Our results show that interlayer moir\'e excitons offer a unique platform for exploring cooperative optical phenomena in strongly interacting many-body systems, thus, holding promise for applications in quantum nonlinear optics.   
\end{abstract}

\maketitle 
Two-dimensional semiconductors such as atomically thin transition-metal dichalcogenides (TMDCs) and their moir\'e heterostructures have recently emerged as a platform to realize strongly correlated electronic phases~\cite{ReganNature2020,XuNature2020,HuangNPhys2021,LiNature2021,lwang2020,SmolenskiNature2021,moiremarvel,KennesMoire}. Excitons in these heterostructures have played a crucial role in unveiling elusive correlated electronic phases~\cite{fci2,ZengNature2023,SpinPolaronMak,LivioKineticMag}. Moreover, the possibility to couple photons and moir\'e superlattices has ushered in broad explorations of photonic and optoelectronic phenomena and applications~\cite{MakNP2016,JinNature2019,TranNature2019,SeylerNature2019,TurunenReview,Zhang2021,Du2023,ZengOptoelectronicsReview}.

However, their optical properties beyond the physics of local light-matter coupling has remained relatively unexplored. While most studies focused on electrostatic interactions between moir\'e excitons \cite{ExcitonMottUCSB,HafeziMott,SufeiMott}, their mutual coupling via the electromagnetic vacuum can result in collective optical effects \cite{lehmberg_radiation_1970, friedberg_frequency_1973,scully_collective_2009,rohlsberger_collective_2010, pellegrino_observation_2014, jennewein_coherent_2016, bromley_collective_2016,huang2024collectiveopticalpropertiesmoire} that are particularly pronounced in the deep subwavelength limit of typical moir\'e lattice spacings. 

In this Letter, we describe how the combination of collective light-matter coupling and strong exciton-exciton interactions gives rise to quantum many-body dynamics and cooperative optics - and most prominently superradiant photon emission \cite{dicke_coherence_1954,gross_superradiance_1982,masson_universality_2022}. More generally, the interplay of strong on-site repulsion, electrostatic dipolar interactions, long-range light-induced excitation hopping, and long-range dissipation suggests moir\'e excitons as an interesting open many-body system. Here, we explore this interplay and its effect on cooperative superradiant emission processes. We reveal pronounced effects of static dipole interactions and show that it suppresses the otherwise strong superradiance of fully excited lattices, while it can generate superradiance by ordered arrays of partially filled lattices that may form due to dipolar repulsion between excitons \cite{ParkDipoleLadder2023}. Electron doping is also found to have remarkably strong effects on cooperative decay, with the formation of generalized Wigner crystals of electrons inducing superradiant emission from otherwise slowly decaying exciton lattices. Our findings highlight the role of static particle interactions in controlling cooperative radiance, and contribute to understanding the optical properties of moir\'e quantum materials as well as their use to probe many-body phenomena of interacting excitons and electrons.

\begin{figure}[t!]
    \centering
    \includegraphics{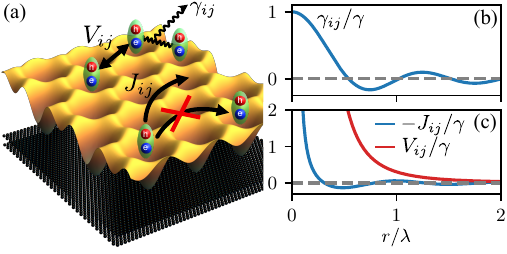}
    \caption{
    (a) Illustration of a moir\'e potential in layered TMDC heterostructures. Interlayer excitons have strong dipolar interactions $V_{ij}$ and undergo long-range photon-mediated hopping with amplitudes $J_{ij}$ unless the site is already occupied by an exciton. Their collective decay is described by distance-dependent rates $\gamma_{ij}$ according to Eqs.(\ref{eq:Ham})-(\ref{eq:JandGamma}), with $\gamma_{ii}=\gamma$, shown in (b). (c) Distance dependence of $J_{ij}$ and $V_{ij}$ for a relative dipolar interaction strength of $\varepsilon_\mathrm{dd} = 5$.
    }
    \label{fig:figure1}
\end{figure}

We consider interlayer excitons formed by an electron-hole pair that are located in separate layers of a two-dimensional heterostructure,  realized by two stacked TMDC monolayers (e.g. WS$_2$/WSe$_2$) as sketched in Fig.~\ref{fig:figure1}(a) \cite{RiveraIX,JinNature2019}. A small lattice mismatch or a finite twist angle between the monolayers generates a slowly varying moir\'e pattern of the potential  for electrons and holes \cite{FengchengHubbard,FengchengMoireExciton}. As the excitons localize in the minima of this potential they form an array with a moir\'e lattice constant $a\sim 10~{\rm nm}$~\cite{ParkDipoleLadder2023}, which is orders of magnitude larger than that of the underlying atomic crystal but lies well below the optical wavelength $\lambda$ of the excitonic transition. As a result, moir\'e lattices constitute a subwavelength array in which exciton-photon interactions can be highly collective, whereby emitted photons can be reabsorbed at other lattice sites. This vacuum-mediated photon exchange generates a long-range coupling between distant moir\'e sites and results in collective photon emission associated with the formation of super- and subradiant states. The deep subwavelength nature of the system, $a\ll\lambda$, results in a collective coupling that can be orders of magnitude larger than the single-site emission rate and is expected to further enhance such collective effects, e.g., compared to neutral atoms in optical lattices~\cite{asenjo-garcia_exponential_2017, shahmoon_cooperative_2017, rui_subradiant_2020, bettles_enhanced_2016, chang_colloquium_2018, srakaew_subwavelength_2023, facchinetti_storing_2016, pedersen_greens_2024,zhang_photon-photon_2022,sierra_dicke_2022,perczel_topological_2017,basler_linear_2023,bekenstein_quantum_2020,moreno-cardoner_quantum_2021,masson_many-body_2020,masson_universality_2022, mink_collective_2023}, where typically $a\lesssim\lambda$.

\begin{figure}
    \centering
    \includegraphics{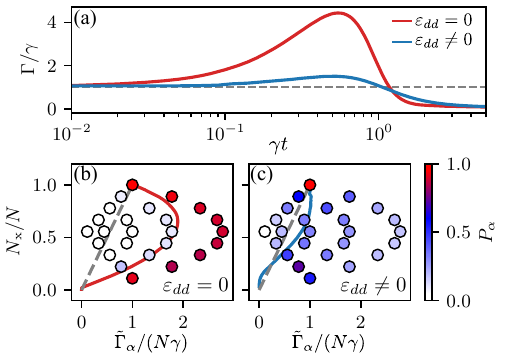}
    \caption{
    (a) The time-dependent cooperative emission rate $\Gamma$ demonstrates strong suppression of superradiance by static dipole-dipole interactions, quantified by the dimensionless interaction strength $\varepsilon_{\rm dd}$. The spectrum of collective decay rates is shown in (b) and (c) for a $3 \times 3$ triangular lattice ($N=9$) with $a/\lambda = 0.05$ and for different numbers $N_{\rm x}$ of excitons in the lattice. States to the right of the dashed line are superradiant, while those to the left are subradiant. The color code shows their the time-integrated population $P_{\alpha}$ during the cooperative decay of an initially fully excited lattice ($N_{\rm x}$), and the solid lines show the time-evolving decay rates of (a). The gray dashed lines correspond to independent photon emission as $N_\mathrm{x} = N \exp(-\gamma t)$.
    }
    \label{fig:figure1_add}
\end{figure}

\begin{figure*}
\includegraphics{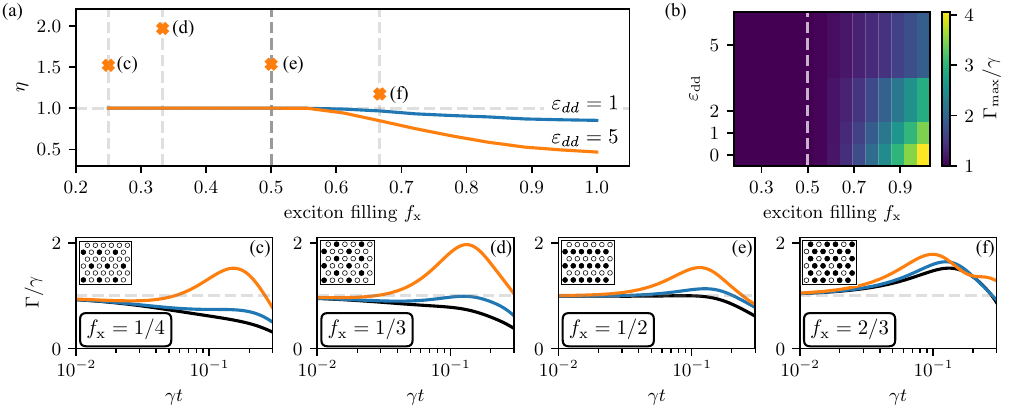}
    \caption{
    (a) Relative enhancement or suppression of the superradiant peak [cf. Eq.(\ref{eq:eta})] due to static dipole-dipole interactions for two interaction strengths $\varepsilon_\mathrm{dd}$ and different initial filling fractions $f_{\rm x}$. The solid lines show results for  random initial configurations. The orange crosses show the enhancement for $\varepsilon_\mathrm{dd} = 5$ and ordered configurations at filling fractions $f_\mathrm{x} = 1/4, \, 1/3, \, 1/2, \, 2/3$, shown in (c)-(f) along with the decay dynamics of these ordered states for $\varepsilon_\mathrm{dd} = 0$ (black line), $\varepsilon_\mathrm{dd} = 1$ (blue line), and $\varepsilon_\mathrm{dd} = 5$ (orange line). (b) The dependence of the superradiant-peak maximum as a function of $f_\mathrm{x}$ and $\varepsilon_\mathrm{dd}$ for random initial configurations.}
    \label{fig:figure2}
\end{figure*}

However, strong exciton-exciton interactions are required to create conditions for cooperative optical phenomena beyond the linear physics of collectively coupled classical dipoles. 
Recently, it was observed that forcing two excitons to occupy the same moir\'e site gives rise to a large Coulomb repulsion energy resulting in an excitonic Mott insulator~\cite{ExcitonMottUCSB,HafeziMott,ParkDipoleLadder2023,SufeiMott}, with a single exciton per moir\'e site. This exclusion of double occupation permits treating interlayer excitons as spin-$1/2$ degrees of freedom, whereby the raising and lowering operators  $\hat{\sigma}^+_i$ and $\hat{\sigma}^-_i$ describe the creation and recombination of an exciton at the $i{\rm th}$ site of the moir\'e lattice. It is this phase-space filling effect at each lattice site that leads to cooperative photon emission as opposed to collective exponential decay of noninteracting excitons.
In addition, the electron-hole separation of the interlayer excitons in different layers implies a sizable static, out-of-plane electric dipole moment, $d$, that is controlled by the distance between the two layers~\cite{LiDipolar,JauregiIX}. For a dipole orientation orthogonal to the 2D layer, this leads to isotropic long-range  interactions 
\begin{equation}
V(\mathbf{r}_i,\mathbf{r}_j)=\frac{d^2}{4\pi \varepsilon_0 \varepsilon_r |\mathbf{r}_i-\mathbf{r}_j|^3}
\end{equation}
between excitons in different ($i \neq j$) moir\'e sites with positions $\mathbf{r}_i$ and $\mathbf{r}_j$. Here, $\varepsilon_r$ is the relative permittivity that accounts for the polarization in the material for a given host medium such as h-BN. Beyond electrostatic two-body interactions, the excitons' oscillating in-plane dipoles couple via vacuum-mediated interactions. The resulting unitary dynamics is determined by the Hamiltonian
\begin{align}\label{eq:Ham}
    \hat{H} =  t \sum_{\langle i , j\rangle}^N \hat{\sigma}_i^+ \hat{\sigma}_j^-+ \sum_{i \neq j}^N \left(\hbar J_{ij} \hat{\sigma}_i^+ \hat{\sigma}_j^- + \frac{V_{ij}}{2} \hat{n}_i \hat{n}_j \right) \, ,
\end{align}
where $\hat{n}_i = \hat{\sigma}_i^+ \hat{\sigma}_i^-$ gives the exciton occupation at site $i$ and $N$ denotes the number of lattice sites. The two-body terms $J_{ij}=J(\mathbf{r}_i - \mathbf{r}_j)$ and $V_{ij} = V(|\mathbf{r}_i - \mathbf{r}_j|)$ describe the photon-mediated hopping of excitons and their static dipole-dipole interaction. The tunneling of excitons between neighboring sites with an amplitude $t$ adds to the photon-induced exchange processes. While $t$ decays exponentially with the lattice constant~\cite{wu_hubbard_2018}, $J(a)$ only decreases algebraically as $J(a)\sim (\lambda/a)^{3}$, such that the value of $a$ controls the relative contribution from each process.
For simplicity, we focus here on conditions where direct exciton tunneling plays a minor role, but an extension to finite values of $t$ is straightforward without further complications. Finally, one can incorporate radiative recombination of the excitons based on the Lindbladian master equation~\cite{lehmberg_radiation_1970}
\begin{align}\label{eq:master}
    \partial_t \hat{\rho} &= i \hbar^{-1}\left[ \hat{\rho}, \hat{H} \right] + \sum_{ij} \gamma_{ij} \left( \hat{\sigma}_j^- \hat{\rho} \hat{\sigma}_i^+ - \frac{1}{2} \left\{\hat{\sigma}_i^+ \hat{\sigma}_j^-, \hat{\rho} \right\} \right) \, 
\end{align}
for the many-body density matrix $\hat{\rho}$ of the system.
Here, the two-body decay terms $\gamma_{ij}$ account for the collective photon emission by the exciton array, whereby the diagonal element $\gamma_{ii} = \gamma$ corresponds to the bare one-body   emission rate due to recombination of a single localized exciton at a given moir\'e site $i$. The corresponding decay rates along with the photon-mediated exchange amplitude are determined as \cite{lehmberg_radiation_1970}
\begin{align}\label{eq:JandGamma}
    J_{ij} - i \frac{\gamma_{ij}}{2} = - \frac{3 \pi \gamma}{\omega} \bar{\mathbf{p}}^\dagger \mathbf{G}(\mathbf{r}_i - \mathbf{r}_j) \bar{\mathbf{p}} \, ,
\end{align}
by the dyadic Green's tensor $\mathbf{G}$ of the electromagnetic field \cite{jackson_classical_2009}, the exciton resonance frequency $\omega=c/(\sqrt{\varepsilon_r}\lambda)$ and the unit vector $\bar{\bf p}={\bf p}/p$ of the transition-dipole moment $\mathbf{p}$. Since the optical transition in TMDC bilayer structures is valley selective, only a single circular polarization is relevant for a valley-polarized initial state of the excitons. Figures \ref{fig:figure1}(b) and (c) show the dependence of $\gamma_{ij}$, $J_{ij}$, and $V_{ij}$ on the distance between excitons. 
The strength of the static dipole-dipole interactions relative to the dipolar exchange of excitons can be quantified by the parameter $\varepsilon_{\rm dd}=2 d^2/p^2$. For small distances $r\ll\lambda$,  the photon-mediated exciton hopping is determined by near-field transition-dipole interactions $J(r) = - 3 \gamma / 8(kr)^3$, such that the parameter $\varepsilon_{\rm dd} \approx {V(a)}/{\hbar \vert J(a) \vert}$ directly gives the ratio of the nearest-neighbor dipole interactions for the typical lattice constants $a\ll \lambda$ considered in this work.

In order to obtain an intuitive picture of the resulting cooperative decay dynamics, we first consider a small triangular lattice with a spacing of $a=\lambda/20$ in the absence of static dipole-dipole interactions. For the chosen parameters and small system size, the decay rates, $\gamma_{ij}\approx \gamma_{ii}=\gamma$ are nearly homogeneous across the lattice. Such a system displays Dicke superradiance above half filling \cite{dicke_coherence_1954,gross_superradiance_1982}, where a short, intense light pulse emerges from the synchronized decay of multiple excitons through stimulated emission, in contrast to the exponentially decaying fluorescence of independent excitons. One can analyze this process using the effective non-Hermitian Hamiltonian 
\begin{equation}\label{eq:nonHermH}
\hat{\mathcal{H}}=-i\sum_{i,j}\gamma_{ij}\hat{\sigma}_{i}^+\hat{\sigma}_{j}^-
\end{equation}
that follows from Eq.(\ref{eq:master}) and its eigenvalues $\tilde{\Gamma}_\alpha$ yield the collective decay rates of the system. We diagonalize Eq. (\ref{eq:nonHermH}) for a small number $N$ of lattice sites in a given subspace of $N_{\rm x}$ excitons. The resulting spectrum of collective decay rates is shown in Figs.\ref{fig:figure1_add}(b) and 2(c). Additionally, we simulate the cooperative decay dynamics of an initially unit filled lattice and determine the time-integrated population $P_\alpha$ of the collective eigenstates with decay rates $\tilde{\Gamma}_\alpha$. For noninteracting excitons [Fig.\ref{fig:figure1_add}(b)], $\varepsilon_{\rm dd}=0$, one observes expectedly that the system predominantly evolves along the most superradiant states as it decays with a rapid burst of emitted photons.

However, as Fig.\ref{fig:figure1_add}(c) illustrates, this behavior is altered significantly by the dipole-dipole interaction $V_{ij}$ between excitons. The caused phase rotations of excitonic Fock states, generate a coupling in the basis of the collective eigenstates of Eq.~(\ref{eq:nonHermH}). As a result, it transfers population out of the most superradiant states toward lower decay rates and thereby suppresses superradiance during photon emission. The effect can also be understood by noting that the most superradiant eigenstate of Eq.~(\ref{eq:nonHermH}), from which all excitons emit in synchrony, is the most symmetric superposition state of delocalized excitons. The dipole-dipole interaction breaks this symmetry for any lattice filling $N_{\rm x}/N<1$ such that the phase fluctuations generated by $V_{ij}$ simultaneously suppress superradiance. 

In order to analyze this effect for larger lattices, we use a third-order cumulant expansion \cite{kubo_generalized_1962, kramer_generalized_2015}, which was previously found to yield a reliable description of superradiant decay dynamics in noninteracting ($\varepsilon_{\rm dd}=0$) systems~\cite{robicheaux_beyond_2021, rubies-bigorda_characterizing_2023, rubies-bigorda_dynamic_2023}. The simulations are performed for lattices with $6\times6$ sites, which  resembles qualitatively the behavior of larger systems and in the thermodynamic limit~\cite{suppl}. The total rate of photon emission is obtained from
\begin{align}
    \Gamma (t) = - \frac{1}{N_\mathrm{x}}\frac{d}{dt}\sum_i \langle \hat{n}_i \rangle = \frac{1}{N_\mathrm{x}}\sum_{i,j} \gamma_{ij} \langle \hat{\sigma}_i^+ \hat{\sigma}_j^- \rangle \, . 
\end{align}
Starting from a state with no spatial coherence, $\langle \hat{\sigma}_i^+ \hat{\sigma}_j^- \rangle=\delta_{ij}\langle \hat{n}_i \rangle$, the initial fluorescence signal simply arises from the independent photon emission of each exciton with a rate $\Gamma(0)=\gamma$ that is given by the sum of the one-body decay rates $\gamma=\gamma_{ii}$ over all $N_{\rm x}=\sum_i\langle\hat{n}_i\rangle$ initially prepared excitons. As the system evolves, the rate changes, leading to either subradiant ($\Gamma<\gamma$) or superradiant ($\Gamma>\gamma$) decay. In order to quantify the effects of the static dipole-dipole interactions, we record the maximum emission rate $\Gamma_{\rm max}$ during the decay dynamics and calculate the relative enhancement or suppression
\begin{align}\label{eq:eta}
    \eta = \frac{\Gamma_{\rm max}(\varepsilon_{\rm dd})}{\Gamma_{\rm max}(\varepsilon_{\rm dd}=0)}
\end{align}
of this rate compared to the noninteracting case, $\varepsilon_{\rm dd}=0$. Figure\ref{fig:figure2} shows this ratio as a function of the initial exciton density, $f_{\rm x}=N_{\rm x}(t=0)/N$, for a randomly filled lattice. In the absence of interactions superradiance sets in for initial filling fractions of $f_{\rm x}\gtrsim 1/2$, and we indeed find a reduction of the maximum rate by dipole-dipole interactions above this value. The suppression of the superradiant burst increases both with the strength $\varepsilon_{\rm dd}$ of the dipole-dipole interaction and with the initial exciton density $f_{\rm x}$. The absolute effect on superradiance is quantified in Fig.~\ref{fig:figure2}(b) showing the maximum emission rate as a function of $\varepsilon_{\rm dd}$ and $f_{\rm x}$. The superradiant burst is strongly diminished by the dipole-dipole interaction and eventually vanishes with increasing interaction strength, even at the maximum initial exciton density of $f_{\rm x}=1$.

Remarkably, this effect can be reversed for ordered initial states, which may be prepared as low-energy states of the interacting dipolar excitons at sufficiently low temperatures. Different examples for specific commensurate lattice fillings are shown in Fig.\ref{fig:figure2}. For initial densities $f_{\rm x}\le 1/2$, the noninteracting decay dynamics is subradiant and predominantly populates collective eigenstates of Eq.(\ref{eq:nonHermH}) with $\tilde{\Gamma}_\alpha<\gamma$ during the decay of the  ordered initial exciton configuration. Consequently, the coupling between the collective eigenstates that is induced by the dipole-dipole interaction can, in this case, transfer population to states with larger decay rates. As demonstrated in Fig.\ref{fig:figure2}(c)-(e) for different ordered configurations, this coupling is indeed strong enough to generate or enhance superradiance. 


\begin{figure}
    \centering
    \includegraphics{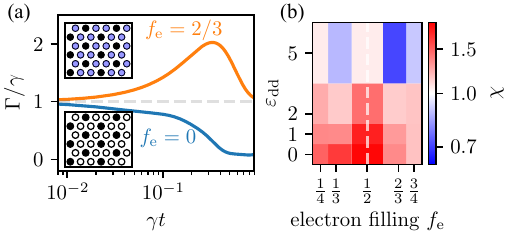}
    \caption{
    (a) Cooperative decay dynamics of an ordered initial state at $f_\mathrm{x} = 1/3$ without interactions ($\varepsilon_\mathrm{dd} = 0$) for no electron doping ($f_\mathrm{e} = 0$, blue curve) and finite electron doping ($f_\mathrm{e} = 2/3$, orange curve). Doping with $ f_\mathrm{e} = 1-f_\mathrm{x}$ can generate superradiance. (b) The interplay of electron doping ($f_{\rm e}$) and excitonic dipole-dipole interactions ($\varepsilon_\mathrm{dd}$) can, however, also suppress cooperative decay (see text for details) The initial exciton filling is $f_\mathrm{x} = 1-f_\mathrm{e}$ in (b).}
    \label{fig:figure3}
\end{figure}

In contrast to atomic systems, it is not straightforward to compare the collective decay dynamics to the one-body decay of isolated emitters, as excitons are typically generated as collective excitations in the semiconducting material. However, exploiting the moir\'e lattice structure this may be possible by  electron doping. Electrically injected charge carriers also localize in the minima of the periodic moir\'e potential forming an insulating phase at unit filling and arranging in generalized Wigner crystals (GWCs) for lower commensurate electron filling fractions $f_{\rm e}$ \cite{ReganNature2020}. In WS$_2$/WSe$_2$ heterostructures, it was observed that the generation of an exciton at the same site as a doped charge carrier is avoided below a total filling $f_{\rm tot} = f_{\rm e} + f_{\rm x} = 1$~\cite{ExcitonMottUCSB,SufeiMott,IntercellMoire}. Thus, electrons and excitons occupy complementary moir\'e sites below $f_{\rm tot} < 1$ and reconfigurable excitonic crystals are expected when electrons form GWCs.  Generating excitons in a doped material just below unit electron filling would quench collective interactions and thus allow preparation of localized and isolated excitons that  individually undergo exponential decay and independently emit photons. Strong electron doping in moir\'e lattices thus offers a way to turn off cooperative radiance and measure the one-body emission rate $\gamma$.

More surprisingly, one finds that electron doping can also generate a superradiant burst. In order to see this, let us consider an ordered initial state of the excitons with a filling fraction $f_{\rm x}=1/3$, as depicted in the lower inset of Fig.~\ref{fig:figure3}(a). As discussed above, such a filling fraction will undergo subradiant decay [see Fig.\ref{fig:figure3}(a)] since the initial exciton density is well below $f_{\rm x}<1/2$. The decay dynamics changes dramatically when doping the system with $f_{\rm e}=2/3$ such that electrons form a GWC as illustrated in the upper inset of Fig.\ref{fig:figure3}(a). While this neither affects the initial density, $f_{\rm x}$, nor the spatial configuration of the excitons, it generates a significant superradiant burst. The effects of doping on the decay dynamics are summarized in Fig.\ref{fig:figure3}(b), where we show the ratio 
\begin{equation}
\chi=\frac{\Gamma_{\rm max}(f_{\rm e}=1-f_{\rm x})}{\Gamma_{\rm max}(f_{\rm e}=0)}
\end{equation}
of the maximum emission rate $\Gamma_{\rm max}$ of a maximally doped lattice and without electron doping. For vanishing dipole-dipole interactions ($\varepsilon_{\rm dd}=0$), electron doping with $f_{\rm e}=1-f_{\rm x}$ consistently increases the superradiant decay rate of the system. Compared to the undoped system, the immobile electrons of the formed Wigner crystal persistently block all unoccupied lattice sites, effectively increasing the excitonic filling fraction of the remaining moir\'e sites to unity and generating a superradiant burst. Interestingly, the effect of doping can even reverse for strong exciton interactions and $f_{\rm x}$, where the static dipole-dipole interactions tend to generate superradiance in the undoped case [cf. Fig.\ref{fig:figure2}(c)-(e)], but strongly suppress superradiance for maximal doping where the effective exciton density is increased to unity. This illustrates that electron doping can induce superradiance even at relatively small exciton filling fractions and modify the emission properties of the system.


Heterobilayer materials have moir\'e lattice constants in the range of $a \sim 10 - 50 \, \mathrm{nm}$ and typical transition energies of $\hbar\omega\sim1\,{\rm eV}$. Their corresponding lattice spacings, therefore, occupy the deep subwavelength regime of $a/\lambda\sim10^{-2}- 10^{-1}$, considered in this work. The radiative  lifetime of moir\'e excitons, which determines the strength $J(r)$ of the photon-mediated exciton hopping, can be as short as $\gamma^{-1}\sim 1\, {\rm ns}$ and as long as $\gamma^{-1}\sim 100 \,{\rm ns}$ \cite{RiveraIX,IXlifetimeMete}. Given typical layer spacings, static dipolar interactions reach values of $V(a)=0.1-1 \,{\rm meV}$, corresponding to interaction strengths of up to $\varepsilon_\mathrm{dd} \sim 10$. Similar lattice constants and static dipole-dipole interactions are found for field-polarized excitons in homobilayer systems, where  periodic potentials are created via a twisted substrate~\cite{datta_highly_2022,cho_moire_2024, han_highly_2024}. Naturally, however, excitons in such settings make it possible to reach stronger light-matter coupling than for interlayer excitons in heterobilayer materials. Dipolar excitons with strong static interactions can also be realized in double quantum wells, where lattice potentials are engineered via  electric fields generated by periodic arrangements of surface electrodes~\cite{togan_enhanced_2018, lagoin_extended_2022, lagoin_evidence_2024}. While this yields larger lattice constants of $a\sim250 \, \mathrm{nm}$, such values remain in the subwavelength regime. While spatial variations of the lattice potential and disorder are generally expected to counteract superradiance in all these settings, spatial homogeneity over only a few lattice sites is sufficient to observe the superradiant decay dynamics discussed in this Letter. Similarly, our simulations for relatively small lattice indicate that inevitable decoherence processes will not diminish the superradiant signal as long as spatial coherence is maintained over a few lattice sites. Indeed, recent experiments on double quantum wells report signatures of cooperative photon emission from 2D lattices of excitons~\cite{lagoin_evidence_2024}, suggesting an exciting outlook for experimental studies of dipolar-interaction effects on superradiance.

In conclusion, the results of this work suggest moiré quantum materials as a promising platform to study many-body cooperative optics of strongly interacting quantum emitters. While we have focused on cooperative decay, exploring the driven-dissipative many-body dynamics and emerging nonequilibrium steady states would present an interesting extension. Here, the doping with electrons at filling fractions away from the Wigner crystalline phases considered in this Letter is appealing for studying driven-dissipative phases of strongly interacting moir\'e electrons and optically pumped moir\'e excitons~\cite{MottDopingExciton}. Moreover, the inclusion of spin-valley degrees of freedom \cite{SchaibleyValleyReview} would permit  optical probing of quantum magnetism in a driven-dissipative setting, whereby the possibility of initializing exciton spins could allow for optical control of electronic spin order~\cite{LightMagnetismMoire}. Given the expectedly correlated nature of the excitonic states that form dynamically during their cooperative decay, understanding the nonclassical statistics and emerging photon correlations in the emitted light present another important question for future work. Effects of static particle interactions on cooperative optical processes, as studied in this Letter, may also become relevant in optical-lattice experiments \cite{browaeys_many-body_2020,kaufman_quantum_2021,rui_subradiant_2020, srakaew_subwavelength_2023} with magnetic atoms \cite{su_dipolar_2023,grun_optical_2024}. 

\begin{acknowledgments}
We thank O. Rubies-Bigorda, S. Yelin, M. Hafezi, A. Imamo\u{g}lu, Z.~J. Hadjri and L.~M. Devenica for insightful discussions. This work was supported by funding from the European Union's Horizon Europe research and innovation program under the Marie Sklodowska-Curie Grant Agreement No. 101106552 (QuLowD), from the Austrian Science Fund (FWF) (Grant DOI: 10.55776/COE1) and the European Union (NextGenerationEU), from the NSF Division of Materials Research (Award No. 1905809), from the State Secretariat for Education, Research and Innovation (SERI)-funded European Research Council Consolidator Grant TuneInt2Quantum (No. 101043957), and from the European Research Council through the ERC Synergy Grant SuperWave (Grant No. 101071882).
\end{acknowledgments}

\bibliography{ReferencesExcitonSuperradiance}

\newpage
\onecolumngrid

\section{Appendix: Finite-size effects}

\begin{figure}[b!]
    \centering
    \includegraphics[width=0.95\linewidth]{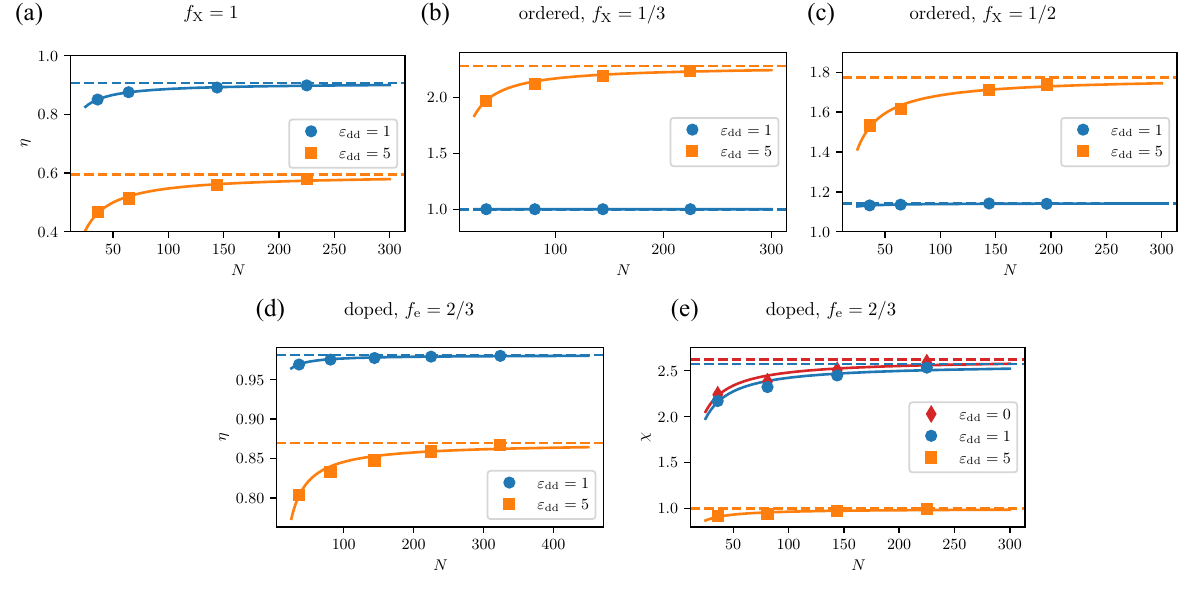}
    \caption{System-size dependence of the reduction and enhancement of the superradiant peak. The symbols show the simulation results, solid lines are a fit to Eq.~(\ref{eq:supp_finite_size}), and dashed lines indicate the asymptotic values as $N \to \infty$. Panel (a)-(c) show the factor $\eta$ for different ordered initial states with indicated lattice fillings $f_X$ and different indicated static dipole-dipole interactions. Panel (d) shows $\eta$ in a doped system with $f_X=1/3$ and $f_e=2/3$, while panel (e) shows the corresponding change, $\chi$, of the superradiant peak due to doping. All other parameters are the same as in Figs.~3 and ~4 of the main text.}
    \label{fig:Figure_supp_finite_size}
\end{figure}

In order to assess the effect of the finite size of the system considered in the main text, we perform a finite-size analysis for selected filling and doping parameters. The results are shown in Fig.~\ref{fig:Figure_supp_finite_size}. Panels (a)-(c) show the  factor $\eta$ as defined in Eq.~(7) of the main text for different ordered initial states with $f_\mathrm{x} = 1$, $1/3$, and $1/2$, respectively. While we observe quantitative differences due to the different system sizes, the general effect of the static dipole-dipole interaction to suppress or enhance superradiance is not affected by finite-size effects. The reduction at unit filling decreases slightly towards larger numbers of lattice sites, whereas the enhancement for ordered states becomes even stronger with increasing system size. Importantly, it can also be seen that the results saturate upon increasing the size of the lattice. Indeed, one expects convergence as the lattice size exceeds significantly the optical wavelength $\lambda$.

Panels (d) and (e) show the relative suppression $\eta$ for a doped system with $f_\mathrm{e} = 2/3$ and $f_\mathrm{x} = 1/3$ and the relative enhancement compared to an undoped system $\chi$ as defined in Eq.~(8) in the main text, respectively. Here, we find qualitatively similar behavior, once again, leaving the conclusions drawn in the main text unaffected by the finite size of the simulated lattices.
We estimate the large-$N$ limit by fitting the simple expression 
\begin{align}
    \eta = \eta_\infty + \frac{\alpha}{N}, \quad N \gg 1\, 
    \label{eq:supp_finite_size}
\end{align}
to the data shown in Fig.~\ref{fig:Figure_supp_finite_size} (and similarly for $\chi$). The fitted curves are shown as solid lines in the figure and the obtained asymptotic values, $\eta_\infty$ and $\chi_\infty$, are indicated by the horizontal dashed lines in Fig.~\ref{fig:Figure_supp_finite_size}.

\end{document}